\documentstyle[11pt,appb,amsmath,epsf]{article}

\begin{document}


\title{Pseudoscalar and vector meson Production in NN Collisions
       \thanks{Invited lecture presented at the symposium on "Threshold Meson 
               Production in $pp$ and $pd$ Interaction", June 2001, Cracow, Poland}
}
\author{K. Nakayama$^{a,b}$
\address{$^a$Department of Physics and Astronomy, University of Georgia, 
Athens, GA 30602, USA \\
$^b$Institut f\"ur Kernphysik, Forschungszentrum-J\"ulich,
D-52425, J\"ulich, Germany}
}

\maketitle

\begin{abstract}
Heavy meson production in nucleon-nucleon collisions is discussed within 
a meson exchange model of hadronic interactions, paying special attention to 
the basic dynamics that determine the behavior of the cross sections near the 
threshold energy. The $pp\rightarrow pp\phi$ reaction is discussed as an 
example of the production of vector mesons in $NN$ collisions.
For the pseudoscalar meson  production, results for the $\eta$ production in 
both the $pp$ and $pn$ collisions are presented.
\end{abstract}

\newpage

\section{ Introduction}

With the advent of particle accelerators in the few $GeV$ energy region, 
heavy meson production in hadronic collisions has attracted increasing 
attention in the past few years. In particular, heavy meson production
in nucleon-nucleon ($NN$) collisions is of special interest because it 
allows to investigate in a simple system the short range hadron dynamics 
for which, so far, we have a very limited knowledge. Due to the large 
momentum transfer between the initial and final states, these reactions 
necessarily probe the dynamics at short distances. In this 
context, apart from the intrinsic interest associated with the particular meson 
produced, the production of pseudoscalar and vector mesons can be used as
a tool to probe the short distance dynamics systematically. Table~\ref{tab1}
illustrates this point: it shows the momentum transfer and the corresponding 
distance probed by producing mesons of different masses at the respective 
threshold energy. At threshold, the momentum transfer is given by 
$q=(m_N m_M + m_M^2/4)^{1/2}$, where $m_N$ and $m_M$ denote the nucleon and meson 
mass, respectively. As one can see, the distance probed in these reactions 
ranges from $0.53 fm$ for pion production to $0.18 fm$ for $\phi$ meson 
production. 
\begin{table} [h]
\caption{Momentum transfer $q$ and the corresponding distance $r$ probed by the 
$NN\rightarrow NNM$ reaction at the threshold energy for different particles $M$ 
produced.}
\vskip -0.25cm
\tabskip=1em plus2em minus.5em
\halign to \hsize{\hfil#&
                  \hfil#\hfil&\hfil#\hfil&\hfil#\hfil\cr
\noalign{\hrulefill}
particle  & mass ($MeV$) & q ($fm^{-1}$) & r ($fm$) \cr
\noalign{\vskip -0.20cm}
\noalign{\hrulefill}
\noalign{\vskip -0.5cm}
\noalign{\hrulefill}
$\gamma$        &   0   &  0.0  &  $\infty$ \cr
$\pi$           & 140   &  1.9  &  0.53 \cr
$\eta$          & 550   &  3.9  &  0.26 \cr
$\rho , \omega$ & 780   &  4.8  &  0.21 \cr
$\eta^\prime$   & 960   &  5.4  &  0.19 \cr
$\phi$          &1020   &  5.6  &  0.18 \cr
\noalign{\hrulefill} }
\label{tab1}
\end{table}

The theory of heavy meson production is still in its early stage
of development. As we have seen above, heavy meson production reactions 
can probe quite short distances - down to less than $0.2 fm$. These short 
distances correspond to the region of confinement where the relevant 
degrees of freedom are the constituent quarks and gluon flux tubes. Thus, the  
appropriate approach to describe these short range dynamics might be the constituent 
quark models rather than the hadronic models. However, such an approach still 
remain to be developed. On the other hand, the transition region from the 
hadronic to constituent quark degrees of freedom does not have a well-defined 
boundary and, consequently, it is of special interest to see how far down 
in distance one can ``push'' the hadronic models. 

In principle, effective field theories can be used to describe near-threshold 
particle production 
processes in terms of hadronic degrees of freedom. However, although such an approach
(Chiral Perturbation Theory) might be appropriate for describing the production of 
pions \cite{ChPT} in spite of the relatively large expansion parameter, 
$Q = \sqrt{m_\pi/m_N} \sim 1/3$, for the description of heavy meson production there 
is, {\it a priori}, no obviously preferred approach. The majority of existing calculations 
of heavy meson production in $NN$ collisions are based on meson exchange models 
of strong interactions \cite{MEM,Sibirtsev,Batinic,Nak1,Gedalin,Nak2,Pena,Wilkin,Baru,MEMEX} 
with a few exceptions \cite{Ahlig}. The price one pays for insisting on such models 
is that their predictions become more and more sensitive to the short range part of the model, 
usually parametrized in terms of the form factors at the hadronic vertices involved. 
The success of such models should be measured in terms of their capability to correlate 
as many independent processes as possible in a consistent manner.
We mention that although the approach used in Ref.\cite{MEMEX} is based on meson exchange 
models, it differs from conventional meson exchange models in a number of aspects, such as the
absence of the form factors at the hadronic vertices, etc.

\section{Meson Exchange Models}

As mentioned in the previous section, the majority of the existing 
calculations of meson production in $NN$ collisions are based on conventional 
meson exchange models. Even within such models, the description of these reactions 
is not a simple task in principle, for the final state is a three-body state and, 
consequently, one needs to solve the three-body Faddeev equation. Of course, a 
complete three-body calculation of this reaction is at present not available 
\cite{TBFadd}. How can one then start to describe these reactions? In order to 
gain some insight to this question, let us examine some of the features exhibited 
by the production cross sections near threshold. In reactions like $NN$ 
bremsstrahlung, where a (massless) photon is produced, the measured cross section
varies with the inverse of the photon energy $\omega_\gamma$ near threshold. This 
feature of the cross section is expressed by the so-called soft-photon theorem 
\cite{Low} which gives
\begin{equation}
\sigma = {A\over \omega_\gamma} + B \ ,
\label{LETg}
\end{equation}
where $A$ and $B$ are constants containing only the on-shell information of the $NN$ 
interaction (or, in other words, the asymptotic behavior of the $NN$ wave function). 
This result is not surprising at all if we recall from Table \ref{tab1} that, close to 
threshold, the photon production reaction in $NN$ collisions probes only the 
asymptotic behavior of the $NN$ wave function. Eq.(\ref{LETg}) is, therefore, regarded 
as a model-independent result and, as such, it should be reproduced by any model 
describing the $NN$ bremsstrahlung reaction.

For production of heavy mesons, we do not have a model independent result like 
the low energy theorem for $NN$ bremsstrahlung. However, as early as 1952, Watson
\cite{Watson} showed that in heavy particle production reactions in which strongly 
(attractively) interacting particles are present, the energy dependence of the cross 
sections should be dictated by the energy dependence of the interaction between those 
strongly interacting particles and the available phase space. In particular, for meson 
production in $NN$ collisions, the energy dependence of the total cross section should 
be given by the energy dependence of the on-shell $NN$ final state interaction (FSI), 
$T(p',p')$, plus the phase space
\begin{eqnarray}
\sigma(E) & \propto & \int d\rho(E,p') |T(p', p')|^2 \nonumber \\
          & \propto & \int d\rho(E,p') \left( {\sin(\delta(p'))\over p'} \right)^2 \ , 
\label{Wat}
\end{eqnarray}
where the integration is over the available phase space, $\rho(E,p')$, with $p'$ denoting 
the relative momentum of the two interacting nucleons in the final state. $\delta(p')$
denotes the corresponding $NN$ phase shift. Watson's result is based on the observation
that the massive particle production is a short range process and, as such, the primary 
production amplitude of such a particle should have a weak energy dependence. Note that the 
production cross section $\sigma(E)$ cannot be expressed in a model-independent way, for 
its absolute value depends on the short range part of the interaction. We shall elaborate 
more on Watson's result later. For the moment, we mention that all the recently measured 
meson production cross sections in $NN$ collisions near threshold follow the energy 
dependence given by Eq.(\ref{Wat}) with the exception of $\eta$ meson production, where one 
sees a relatively small deviation at energies close to threshold. This deviation is commonly 
attributed to the $\eta N$ FSI.

\begin{figure}[h]
\begin{center}
\leavevmode
\epsfxsize=10.0cm
\epsfysize=7.0cm
\epsfverbosetrue
\epsffile{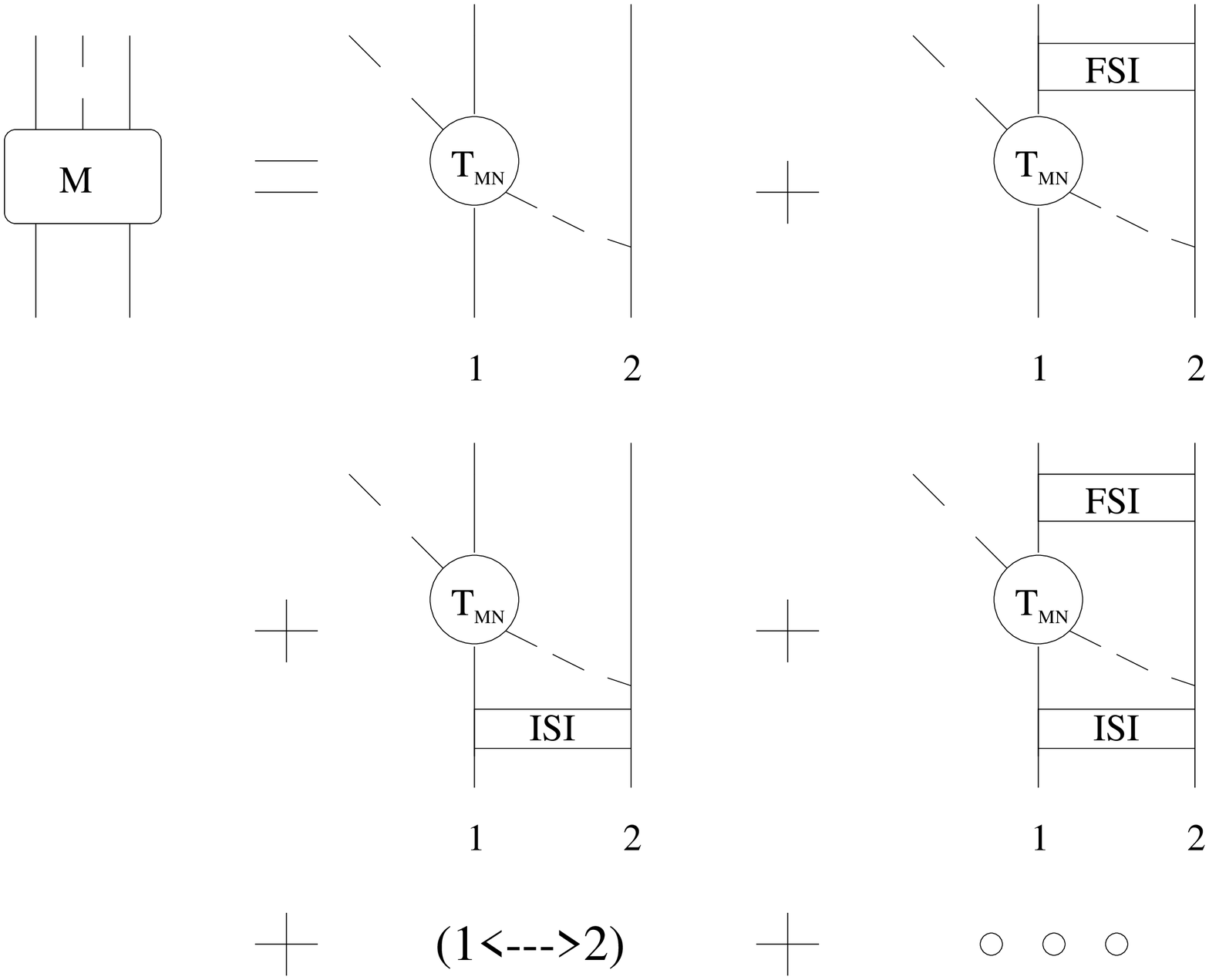}
\end{center}
\caption{Meson production amplitude obtained in a diagrammatic
approach to the $NN\rightarrow NNM$ reaction considering the
$MN$ and $NN$ $T$-matrices as the basic building blocks. $T_{MN}$ stands for
the $MN$ $T$-matrix. FSI and ISI stand for the final and initial state $NN$ 
$T$-matrices, respectively. The first diagram on the r.h.s. is referred to 
as the production current $J$ which enters in other diagrams as can be noted.}
\label{fig0}
\end{figure}

The above consideration indicates that any model of heavy meson production reaction in $NN$
collisions should have built in the feature given by Watson's result, Eq.(\ref{Wat}).
As we shall show in the following, this can be achieved within a Distorted Wave Born
Approximation. Here we follow a diagrammatic approach to derive the total amplitude. 
We start by considering the meson-nucleon ($MN$) and $NN$ interactions 
as the building blocks for constructing the total amplitude describing the $NN\rightarrow
NNM$ reaction. We then consider all possible combinations of these building blocks in a 
topologically distinct way, with two nucleons in the initial state and two nucleons plus a 
meson in the final state. In this process of constructing the total amplitude, care must 
be taken in order to avoid diagrams that lead to double counting. Specifically, these 
are the diagrams that lead to the mass and vertex renormalizations, since we choose to use
the physical masses and coupling constants. The resulting amplitude constructed in this 
way is displayed in Fig.~\ref{fig0}. 
The ellipsis indicates those diagrams that are more 
involved numerically (including, in particular, the $MN$ FSI, which otherwise would be 
generated by solving the three-body Faddeev equation). So far there are very few attempts 
to account for them \cite{Gedalin,Moalem}. Therefore, neglecting those diagrams,
the total amplitude is given by the diagrams displayed explicitly in Fig.~\ref{fig0} and  
reads
\begin{equation}
M = (1 + T^{(-)\dagger}_fG^{(-)*}_f) J (1 + G^{(+)}_iT^{(+)}_i) \ ,
\label{ampl0}
\end{equation}
where $T_{(i,f)}$ denotes the $NN$ $T$-matrix interaction in the initial($i$)/final($f$) state 
and, $G_{(i,f)}$, the corresponding two nucleon propagator. The superscript $\pm$ in 
$T_{(i,f)}$ as well as in $G_{(i,f)}$ indicates the boundary condition ($(-)$ for incoming and 
$(+)$ for outgoing waves). The production current is denoted by $J$, which is nothing other than 
the $MN$ $T$-matrix, $T_{MN}$, with one of the meson legs attached to a nucleon (first diagram
on the r.h.s. in Fig.~\ref{fig0}). Eq.(\ref{ampl0}) is the basic formula on which the large 
majority of existing calculations are based.

We now wish to exhibit the essential features of the meson production reaction contained in
Eq.(\ref{ampl0}). In particular, we would like to make close contact between the 
amplitude $M$ and Watson's result given by Eq.(\ref{Wat}). To this end, we use the 
relation
\begin{equation}
G^{(\pm)}_\alpha = {\cal P}\left( {1\over E_\alpha - E(k)} \right) 
         \mp i\pi\delta(E_\alpha - E(k)) \ ,
\label{prop}
\end{equation}
where $\alpha=i,f$, to express $M$ as
\begin{eqnarray}
M & = &   \left\{1 - i\kappa_f T_f(p', p')
                 [1 + i{1\over ap'}{\cal P}_f(E,p')] \right\} 
   J(E,p') \nonumber \\
  & \times &  \left\{1 - i\kappa_i T_i(p, p)
                 [1 + i{1\over ap}{\cal P}_i(E,p')] \right\}  \ ,
\label{ampl1}
\end{eqnarray}
where $T_{i}(p,p)$ denotes the on-shell $NN$ T-matrix with the relative momentum
$p$ of the interacting two nucleons in the initial state $i$ and $T_f(p', p')$
denotes the on-shell $NN$ T-matrix with the relative momentum $p'$ of the 
interacting two nucleons in the final state $f$. Here the superscript $(+)$ of 
$T_{(i,f)}$ has been omitted. $\kappa_i\equiv \pi p \varepsilon(p) / 2$ 
and $\kappa_f \equiv \pi p' \varepsilon(p') / 2$ are the phase space densities of 
the two nucleons in the initial and final states, respectively. $\varepsilon(q)
\equiv \sqrt{q^2+m_N^2}$ and $a$ is a constant (scattering length) introduced for 
convenience. Note that in the above equation, only those arguments 
of the quantities relevant to the discussion related to Watson's result are 
explicitly displayed. $E \equiv E_i$. The function ${\cal P}_i$ is given by 
\begin{subequations}
\begin{eqnarray}
{\cal P}_i(E, p') & = & \left({ap\over \kappa_i}\right)
                    {\cal P} \int^\infty_0 dk k^2{f_i(E,k)\over E_i - E(k)} \ , \\
\label{P_ia}
f_i(E,k) & \equiv & {T_i(p, k) J(E, k)\over T_i(p, p) J(E, p')} 
                =   {K_i(p, k) J(E, k)\over K_i(p, p) J(E, p')}\ .
\label{P_ib}
\end{eqnarray}
\label{P_i}
\end{subequations}
Similarly,
\begin{subequations}
\begin{eqnarray}
{\cal P}_f(E, p') & = & \left({ap'\over \kappa_f}\right)
                    {\cal P} \int^\infty_0 dk k^2{f_f(E,k)\over E_f - E(k)} \ , \\
\label{P_fa}
f_f(E,k) & \equiv & {T_f(p', k) A(E, k)\over T_f(p', p') A(E, p')}
               =    {K_f(p', k) A(E, k)\over K_f(p', p') A(E, p')}\ ,
\label{P_fb}
\end{eqnarray}
\label{P_f}
\end{subequations}
with $K_{(i,f)}$ denoting the $NN$ K-matrix and $A \equiv (1 + G_iT_i) J$.
The functions ${\cal P}_{(i,f)}(E,p')$ summarize all the off-shell effects of the $NN$ 
interaction and production current. As such, they are unmeasurable and model-dependent 
quantities. For production of heavy mesons near threshold, the off-shell $NN$ interaction
required in ${\cal P}_f(E,p')$ is at very low energy. In calculations based on 
conventional meson exchange models this function is very large and cannot be neglected. In 
contrast, the function $(1/ap){\cal P}_i(E,p')$ is relatively small. This is because for 
heavy meson production the energy of the two nucleons in the initial state has to be large 
enough to produce the meson in the final state. For example, for $\eta$ meson production, 
the incident energy of the beam nucleon corresponding to the threshold energy is about 
$1.25\ GeV$. At such high energies, the on-shell $NN$ interaction has a rather weak energy 
dependence. Note that the inverse scattering theory tells us that, for a local and 
energy-independent $NN$ potential, the off-shell behavior of the $NN$ amplitude is 
completely determined if one knows the corresponding on-shell amplitude in the entire 
energy domain. At least in the case of meson exchange models, the weak energy dependence of
the on-shell $NN$ interaction implies also a flat off-shell behavior and leads to a small 
value of $(1/ap){\cal P}_i(E,p')$. Therefore, at least for the discussion of the essential 
features of the cross section near threshold, the function $(1/ap){\cal P}_i(E,p')$ may be 
neglected.

Using then the relation between the on-shell T-matrix and the phase shift $\delta(q)$ 
and inelasticity $\eta (q)$
\begin{equation}
\kappa(q) T(q,q) \ =  \ \frac{i}{2}\left( \eta (q) e^{2i\delta(q)}-1\right) \ ,
\label{phsft}
\end{equation}
Eq.(\ref{ampl1}) can be reduced to
\begin{eqnarray}
M & \cong &  \left\{ -e^{i\delta_f(p')} \left({\sin(\delta_f(p'))\over ap'} \right)
                 [1 + {\cal P}_f(E,p')] \right\}
        J(E,p') \nonumber \\
  & \times &  \left\{ {1\over 2} \left(\eta_i(p) e^{i2\delta_i(p)} + 1 \right)  
                     \right\} \ ,       
\label{ampl2}
\end{eqnarray}
where the inelasticity in the final $NN$ state is set to unity, $\eta_f(p') = 1$.
The above equation is the desired result. It exhibits the essential features of the
meson production reaction in $NN$ collisions. First, it shows that the relevant energy 
dependence of the total amplitude near threshold is indeed determined by the strong 
energy dependence of the on-shell $NN$ FSI, proportional to $\sin(\delta_f(p'))/ap'$. 
This, when combined with the phase space factor, determines the energy dependence of the 
cross section as given by Eq.(\ref{Wat}). Note that for heavy meson production the 
production current $J(E,p')$, as well as ${\cal P}_f(E,p')$, should be weakly 
energy-dependent, for they summarize all the short range dynamics of the reaction. Thus, 
none of the terms in Eq.(\ref{ampl2}), apart from the on-shell $NN$ FSI, should introduce 
any significant energy dependence near threshold; they just amount to a constant. 
Therefore, the total cross section data near threshold determine just this constant. 
Second, the major effect of the $NN$ initial state interaction (ISI) (the term in the 
last curly brackets) is to reduce the cross section by a factor of      
\begin{eqnarray}
\nonumber
\lambda & = & \left| \frac{1}{2} \left(\eta_i(p) e^{i2\delta_i(p)} + 1 \right) \right|^2
 \\     & = & \eta_{i}(p) \cos ^2(\delta_{i}(p)) + 
              \frac{1}{4}[1-\eta_{i}(p)]^2 \leq \frac{1}{4}[1+\eta_{i}(p)]^2 \ .
\label{isieff}
\end{eqnarray}

As mentioned before, the majority of the existing calculations of meson 
production in $NN$ collisions are based on Eq.(\ref{ampl0}). The differences among 
them reside in how the production current $J$ is modeled, as well as in the   
different treatments of both the $NN$ FSI and ISI. Let's first concentrate on the $NN$
interactions. It is clear that the description of meson production processes in $NN$ 
collisions based on Eq.(\ref{ampl0}) requires the half-off-shell $NN$ ISI and FSI. 
This has been exhibited explicitly in Eq.(\ref{ampl1}) through the functions 
${\cal P}_{(i,f)}$ given by Eqs.(\ref{P_i},\ref{P_f}). Although 
the on-shell $NN$ interaction can be determined from the $NN$ elastic scattering 
experiments, the off-shell behavior of it can only be provided by a given model of the 
$NN$ interaction. For energies below pion threshold, there exist a number of 
accurate meson exchange models \cite{MHE87,Lacomb,Nijmegen} which can provide the 
half-off-shell extension of the $NN$ interaction. As previously noted, these 
so-called realistic $NN$ interactions, based on meson exchange models, yield a rather 
large value for the function ${\cal P}_f(E,p')$  that, consequently,
cannot be neglected. In this connection, for production of heavy mesons, the predicted 
total cross sections can easily differ by a factor of two or more due to the 
different off-shell behavior of these realistic $NN$ FSIs. This indicates that a 
consistent treatment of the $NN$ FSI and the production current $J$ is required. 
Maintaining the consistency between the $NN$ FSI and the production current, however, 
is not a trivial task. While in models where the underlying meson exchange structure is 
known this consistency can, in principle, be maintained, in the case of purely phenomenological 
models, such as the parametrized version of the Paris $NN$ interaction \cite{Lacomb}, such a 
consistency is impossible to achieve -  even in principle. In the excess energy region 
below $Q \cong 100\ MeV$ however, the introduced difference in the predicted total cross 
section due to off-shell differences of these realistic $NN$ interactions is practically a 
constant. It should also be mentioned that the procedure of just evaluating $J$ in the 
on-shell tree level approximation and simply multiplying it by the on-shell $NN$ FSI,
as has been done by many authors (see references quoted in \cite{Hanhart}), is not 
acceptable for obtaining quantitative predictions. As it can be seen from Eq.(\ref{ampl2}), 
the strength of the amplitude $M$ depends on the function ${\cal P}_f(E,p')$.  Using just 
the on-shell $T$-matrix as the FSI instead of the full half-off-shell $T$-matrix means 
setting the function ${\cal P}_f(E,p')$ to zero. Such a procedure simply lacks a consistent 
regularization scheme between the $NN$ interaction and the production current $J$ 
\cite{Hanhart}.

One of the major limitations in the current models of heavy meson 
productions in $NN$ collisions is the lack of a reliable model for the $NN$ ISI. As 
mentioned before, the energy of the two interacting nucleons in the initial state must 
be large enough to produce a meson. For example, for the $\eta$ meson this means nucleon 
incident energies of at least $1.25\ GeV$. For heavier mesons, like $\eta '$ and 
$\phi$, the corresponding threshold incident energy is well above $2\ GeV$, where 
no reliable model exists to provide the half-off-shell $NN$ $T$-matrix that is required
in the evaluation of the function ${\cal P}_i(E,p')$ in Eq.(\ref{P_i}). Note that,
although this quantity is expected to be small and may be neglected for estimates of 
cross sections, for more quantitative predictions it should be taken into account. 
In particular, predictions of spin observables are expected to be sensitive to this
function. The situation with the $NN$ ISI is even worse in the case of heavy meson 
production in $pn$ collisions. There, we lack even the on-shell $NN$ interaction.
While for total isospin $T=1$ states rather reliable phase shift analyses exist up to 
$3\ GeV$ incident energy, for $T=0$ states the reliability is limited to $1.3\ GeV$ 
\cite{SAID,Igor}. This situation imposes a severe limit in all 
existing models of production of mesons heavier than the $\eta$ meson in $pn$ collisions.

There are basically three different approaches in literature in modeling the production 
current $J$ based on meson exchange models. One is a microscopic model of $MN\rightarrow
M'N'$ reactions whose data are reproduced by the model \cite{Batinic,Baru}. The off-shell 
behavior of the $MN$ 
$T$-matrices that are required in the construction of $J$ is then provided by the model. 
Another approach is a ``pseudo'' empirical model in which the production current is 
constructed using the on-shell amplitudes of the $MN\rightarrow M'N'$ as well as 
$\gamma N \rightarrow M'N'$ reactions extracted directly from the available data 
\cite{Sibirtsev,Wilkin}. In the latter reaction the VMD is used to convert its on-shell 
amplitude to an $MN\rightarrow M'N'$ amplitude ($M=\rho,\omega)$. The off-shell behavior of 
the corresponding amplitudes necessary to construct $J$ is then an assumption in this approach. 
The third approach is to split the $MN$ $T$-matrix into the pole ($T^P$) and non-pole ($T^{NP}$) 
parts and consider the non-pole part in the Born approximation only in order to construct the 
production current $J$ \cite{Gedalin,Nak2}. As has been shown elsewhere \cite{Pearce}, the $MN$ 
$T$-matrix can be split into the pole and non-pole parts according to
\begin{equation}
T = T^P + T^{NP} \ ,
\label{MNT}
\end{equation}
where
\begin{equation}
T^P = \sum_B f_{MNB}^\dagger g_B f_{MNB} \ ,
\label{MNTP}
\end{equation}
with $f_{MNB}$ and $g_B$ denoting the physical meson-nucleon-baryon ($MNB$) vertex and baryon
propagator, respectively. The summation runs over the relevant baryons $B$. The non-pole part
of the $T$-matrix is given by
\begin{equation}
T^{NP} = V^{NP} + V^{NP}GT^{NP} \ ,
\label{MNTNP}
\end{equation}
where $V^{NP}\equiv V - V^P$, with $V^P$ denoting the pole part of the $MN$ potential $V$.  
$V^P$ is given by the equation analogous to Eq.(\ref{MNTP}) with the renormalized vertices 
and propagators replaced by the corresponding bare vertices and propagators.
This third approach then leads to a production current which is obtained by approximating the 
full $MN$ $T$-matrix as $T \cong T^P + V^{NP}$.

\section{Our Model}
\begin{figure}[h]
\begin{center}
\leavevmode
\epsfxsize=10.0cm
\epsfysize=7.0cm
\epsfverbosetrue
\epsffile{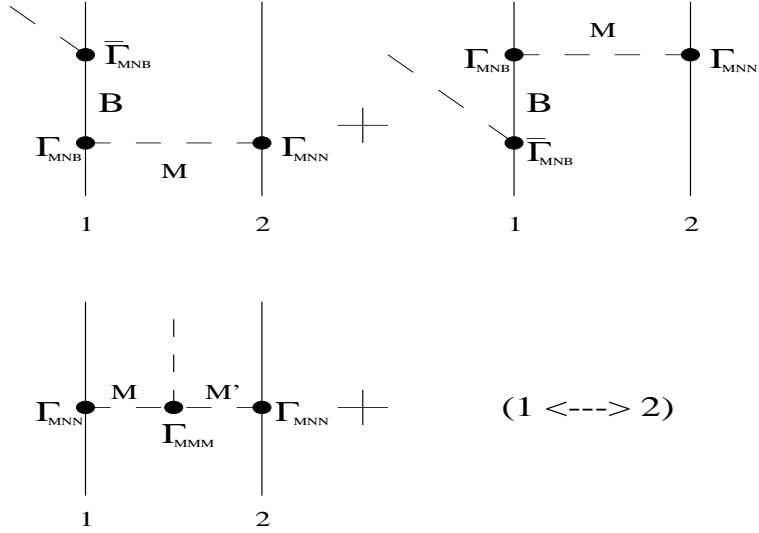}
\end{center}
\caption{Model for the meson production current $J$. The
upper two diagrams are called nucleonic or resonance current depending on the 
intermediate baryon $B$ being a nucleon ($N$) or a nucleon resonance ($N^*$).
The lower diagram is the mesonic current.
$M,M' = \pi, \eta, \rho, \omega, \sigma, a_o$.}
\label{fig1}
\end{figure}

In our model of $NN\rightarrow NNM$, the reaction amplitude $M$ is calculated using
Eq.(\ref{ampl0}). It is based on a relativistic meson exchange model of hadronic 
interactions. The results we shall show in the next section are obtained by using the
$NN$ interaction developed by the Bonn group \cite{MHE87}. This interaction is obtained
by solving a three dimensionally reduced (Blankenbecler-Sugar) version of the
Bethe-Salpeter equation. The loop integrals in Eq.(\ref{ampl0}) are calculated
consistently with the three dimensional reduction 
used in constructing the $NN$ interaction, i.e., the two-nucleon propagators $G_{(i,f)}$
are taken to be the Blankenbecler-Sugar propagator. The production current $J$ is 
modeled according to the third approach discussed in the previous section. It consists of the 
nucleonic, resonance and mesonic currents as displayed diagrammatically in Fig.~\ref{fig1}. 
Note that they are all Feynman diagrams and, as such, they include both the positive- and 
negative-energy propagation of the intermediate particles. 

All the parameters of our model for the production current (coupling constants and form 
factors) are confined to the hadronic vertices $\Gamma_{MNB}$, $\bar\Gamma_{MNB}$ and
$\Gamma_{MMM}$ indicated in Fig.~\ref{fig1} ($B = N,N^*$). Below, we discuss briefly these
parameters for each vertex. For more details, see Refs.\cite{Nak2,Nak3}. 
\begin{itemize}  
\item[$\bar\Gamma_{MNB}$:]
At the meson production vertex $\bar\Gamma_{MNN}$, the coupling constant is taken 
consistently with that used in the construction of the $NN$ interaction. In addition, 
this vertex contains an extra form factor ($F_N(p^2)$) to account for the off-shellness of 
the intermediate nucleon, which is far off-shell if a heavy meson is produced. It is given by
\begin{equation}
F_N(p^2) = {\Lambda_N^4 \over \Lambda_N^4 + (p^2 - m_N^2)^2} \ ,
\label{NFF}
\end{equation}
with $\Lambda_N=1.2\ GeV$. $p$ and $m_N$ stand for the four-momentum and mass 
of the intermediate nucleon, respectively.      

The coupling constant in the vertex $\bar\Gamma_{MNN^*}$ is extracted from the measured 
decay width of the resonance into a meson and a nucleon, $N^*\rightarrow M + N$, whenever
available \cite{PDG}. For the vertex involving a vector meson, the coupling constant
can be extracted from the measured radiative decay using the VMD. The sign of the 
coupling constant is chosen in accordance with the relevant photo-production reaction
analysis \cite{Benmer,RCOUP}. $\bar\Gamma_{MNN^*}$ is also multiplied by the 
form factor given by Eq.(\ref{NFF}), with $m_N$ replaced by $m_{N^*}$.
\item[$\Gamma_{MNB}$:] 
The vertex $\Gamma_{MNN}$ is taken consistently with that used in the construction 
of the $NN$ interaction. For $\Gamma_{MNN}$ involving the off-shell nucleon that
produces the meson, it is also multiplied by the form factor $F_N(p^2)$ given by 
Eq.(\ref{NFF}). The vertex $\Gamma_{MNN^*}$ is the same as $\bar\Gamma_{MNN^*}$, except
that it contains an extra form factor due to the off-shell meson which is taken to be the 
same as that of the corresponding vertex $\Gamma_{MNN}$.
\item[$\Gamma_{MMM}$:]
The coupling constants in the three-meson vertices ($\Gamma_{MMM}$) are extracted from both 
the strong and radiative (measured) decay widths \cite{PDG} in combination with SU(3) 
symmetry, plus the OZI rule \cite{OZI}. The latter relates the basic SU(3) octet and singlet 
coupling constants. $\Gamma_{MMM}$ includes the form factor
\begin{equation}
F_M(q^2,q{'}^2) = \left( {\Lambda_M^2 - m_M^2 \over \Lambda_M^2 - q^2} \right) 
             \left( {\Lambda_{M'}^2 - x m_{M'}^2 \over \Lambda_{M'}^2 - q{'}^2} \right) \ , 
\label{MFF}
\end{equation}
with $\Lambda_M = \Lambda_{M'} = 1.45\ GeV$. $q$ and $q'$ denote the four momenta of the 
two mesons with masses $m_M$ and $m_{M'}$, respectively, that fuse to produce the third 
meson. The parameter $x (=0,1)$ ensures the proper normalization according to the 
normalization point $q{'}^2 = 0, m_{M'}^2$ at which the corresponding coupling constant is
extracted . 
\end{itemize}
There is a number of features in the model described above which are perhaps worth mentioning
here. This model is suited for a systematic analysis of the production of different 
mesons within the same model due to the simplicity of modeling the production current. 
Also, the model is especially suited for studying the role of different reaction mechanisms 
that produce a meson. The consistency between the $NN$ interaction and the production current 
can be easily maintained when one knows the underlying meson exchange structure of the
$NN$ interaction used (note that the way the parameters of our model are fixed ensures the 
consistency between the $NN$ interaction and the production current). As mentioned before, 
this is critical if one wishes to achieve quantitative predictions, especially for production 
of heavy mesons.

\section{Some selected results}
In this section we shall present some selected results on the vector and pseudoscalar meson  
productions based on the model described in the previous section. As an example of
vector meson production, we consider $\phi$ meson production in $pp$ collisions. For 
the pseudoscalar meson production, we discuss the results for $\eta$ production in both the 
$pp$ and $pn$ collisions.

\subsection{$pp\rightarrow pp\phi$}
 
The study of this reaction is of particular importance in connection with the questions 
related to the amount of hidden strangeness in 
the nucleon\cite{Don86EMC88,Ell89Hen92}. In the context of $\phi$ meson production 
processes one expects \cite{Ell89Hen92} that a large amount of hidden strangeness 
in the nucleon would manifest itself in reaction cross sections that significantly 
exceed the values estimated from the OZI rule \cite{OZI}. This phenomenological rule 
states that reactions involving disconnected quark lines are forbidden. In the
naive quark model the nucleon has no $\bar ss$ content, whereas the 
$\phi$ meson is an ideally mixed pure $\bar ss$ state. Thus, in this case, the
OZI rule implies vanishing $\phi NN$ coupling and, 
accordingly, a negligibly small production of $\phi$ mesons from nucleons by 
non-strange hadronic probes.  In 
practice there is a slight deviation from ideal mixing of the vector mesons, 
which means that the $\phi$ meson has a small $\bar uu + \bar dd$ component. 
Thus, even if the OZI rule is strictly enforced, there is a non-zero coupling
of the $\phi$ to the nucleon, although the coupling is very small. Its value 
can be used to calculate lower limits for corresponding cross sections. For 
example, under kinematic conditions chosen to cancel out phase space effects, 
one expects cross section ratios of reactions involving the production of a 
$\phi$- and an $\omega$ meson, respectively, to be
\begin{equation}
R = {\sigma (A+B\rightarrow \phi X)
  \over \sigma (A+B\rightarrow \omega X)}  \approx \tan ^2 (\alpha_V)\ ,
\label{Ratio}
\end{equation}
where $A$, $B$ and $X$ are systems that do not contain strange quarks. 
$\alpha_V \equiv \theta_V - \theta_{V(ideal)}$ is the deviation from the ideal 
$\omega - \phi$ mixing angle. This result arises from simply equating the cross 
section ratio to the square of the ratio of the relevant coupling constants 
at the $\phi$ and $\omega$ production vertices. According to the OZI rule plus 
SU(3) symmetry, these couplings are proportional to $\sin(\alpha_V)$ and 
$\cos(\alpha_V)$, respectively. With the value $\alpha_V \cong 3.7^o$ \cite{PDG}
one gets a rather small ratio of $R = 4.2 \times 10^{-3}$.
The data presented by the DISTO collaboration \cite{Disto1,Disto2} in $pp$ collisions 
indicate that this ratio, after correcting for phase space effects, is about eight times 
larger than the above OZI estimate.
 
In principle, the $\phi$ meson production cross section can be used for a direct 
determination of the $\phi NN$ coupling strength. Any appreciable $\phi NN$ coupling in 
excess of the value given by the OZI rule ($g_{\phi NN} \cong -(0.60 \pm 0.15)$) 
\cite{Nak2} could be interpreted as evidence for hidden strangeness in the nucleon. Of 
course, there is also an alternative picture: one in which the coupling of the $\phi$ meson 
to the nucleon does not occur via possible $\bar ss$ components in the nucleon, but via 
intermediate states with strangeness. Specifically, this means that the $\phi$ meson couples 
to the nucleon via virtual $\Lambda K$, $\Sigma K$, etc. states. Corresponding model 
calculations \cite{Geiger,Meissner} have shown, however, that such processes give rise to 
(effective) $\phi NN$ coupling constants comparable to the OZI values and therefore 
should not play a role in drawing conclusions concerning hidden strangeness in the nucleon. 

\begin{figure}[h]
\begin{center}
\leavevmode
\epsfxsize=10.0cm
\epsfysize=7.0cm
\epsfverbosetrue
\epsffile{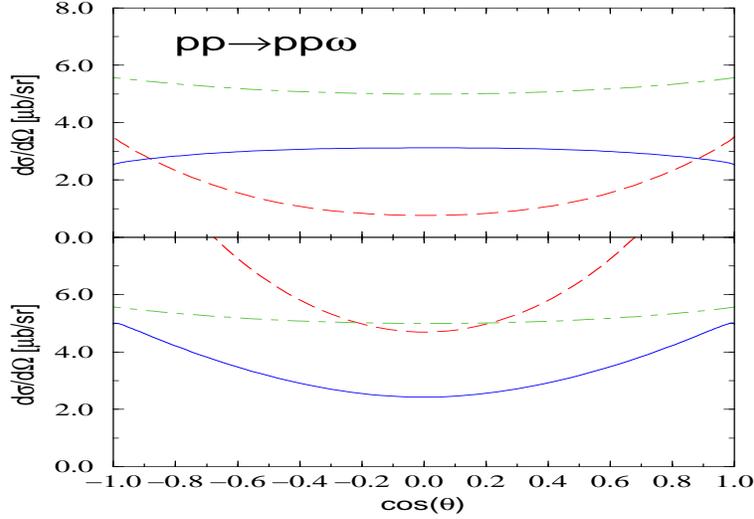}
\end{center}
\caption{Angular distribution of the emitted $\omega$ meson in the c.m. frame of the total
system for two possible scenarios at a proton incident energy of $2.2\ GeV$. The upper panel 
corresponds to the case when the
nucleonic current is smaller than the mesonic current. The lower panel corresponds to
the case when the nucleonic current is larger than the mesonic current. The dashed curves
represent the nucleonic current contribution while the dash-dotted curves represent the
mesonic current contribution. The solid curves correspond to the total contribution.}
\label{fig2}
\end{figure}

Details of our calculation may be found in Ref.\cite{Nak2}. The work of Ref.\cite{Nak2} is 
a combined analysis of the $\omega$ and $\phi$ meson production in $pp$ collisions in 
order to reduce the number of free parameters in the model. Here we only report on the 
essential results. For the production current $J$ we consider only the nucleonic and mesonic 
currents. The latter consists of $v\rho\pi$ exchange current ($v=\omega,\phi$). Other 
mesonic currents have been investigated in a systematic way and found to be very small. 
The nucleon resonance current is not considered because, presently, there is no established 
resonance that decays into a vector meson and a nucleon (see Ref.\cite{Oh} in this connection). 
It is found that the $\phi$ meson is produced significantly from the $\phi\rho\pi$ exchange 
current in $pp$ collisions. This means that one must be able to disentangle this current 
from the nucleonic current (where the $\phi$ meson is emitted directly from the nucleon) 
if one wishes to extract the $\phi NN$ coupling strength from this reaction. As pointed out 
in Ref.\cite{Nak1}, this can be done by looking at the angular distribution of the emitted 
meson. Because the mesonic current contribution yields an isotropic angular distribution, whereas 
the nucleonic current contribution exhibits a $\cos^2(\theta)$ dependence, the two currents 
can be disentangled uniquely from the angular distribution data. Fig.~\ref{fig2} illustrates
this point for the $pp \rightarrow pp\omega$ reaction at a proton beam energy of $2.2\ GeV$.
The upper panel illustrates the situation when the nucleonic current (dashed curve) is smaller 
than the mesonic current (dash-dotted curve) resulting in a nearly flat angular distribution 
(solid curve). Note that the interference between the two currents is destructive, which is
a direct consequence of the signs of the relevant coupling constants determined in our model as 
discussed 
in the previous section. The lower panel shows the situation when the nucleonic current is 
larger than the mesonic current. Here the angular distribution exhibits a pronounced angular 
dependence. In both scenarios the total cross sections have been kept to be about the same.

\begin{figure}[h]
\begin{center}
\leavevmode
\epsfxsize=10.0cm
\epsfysize=7.0cm
\epsfverbosetrue
\epsffile{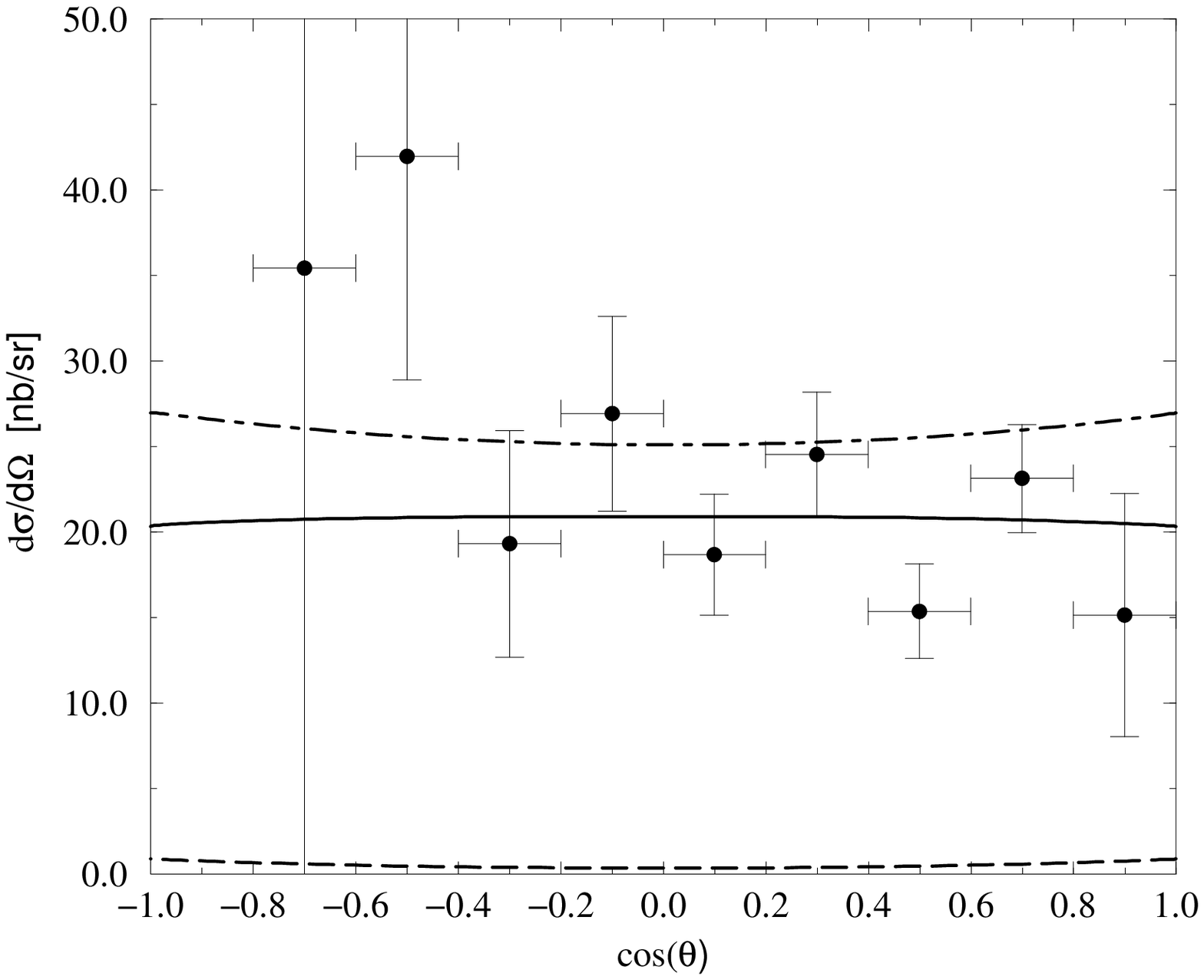}
\end{center}
\caption{Angular distribution of the emitted $\phi$ meson in the c.m. frame of the total
system at a proton incident energy of $2.85\ GeV$. The dashed curve corresponds to
the nucleonic current contribution while the dash-dotted curve to the mesonic current 
contribution. The solid curve corresponds to the total contribution. The data are from
Ref.\protect\cite{Disto1}. The absolute normalization of the data has been determined as 
described in Ref.\protect\cite{Nak2}.}
\label{fig3}
\end{figure}

The earlier data of the $\phi$ meson angular distribution by the DISTO collaboration 
\cite{Disto1} at a proton beam energy of $2.85\ GeV$ is shown in Fig.~\ref{fig3} together with 
our result. The absolute normalization of this data has been determined as described in 
Ref.\cite{Nak2}. Although the data have large uncertainties, the observed angular distribution
is rather flat. Note that the angular distribution should be symmetric about $\theta=90^o$, due
to the identity of the two protons. 
The solid curve is one of our calculations fitted to the 
data. The dash-dotted and dashed curves are the corresponding mesonic and nucleonic current 
contributions, respectively. As one can see, the data require a very small contribution from 
the nucleonic current. The value of the $\phi NN$ coupling constant thus extracted is in the 
range $g_{\phi NN} \approx - (0.2 - 0.9)$. More data for both the $\phi$ and $\omega$ meson 
at low excess energies are necessary in order to determine better the parameters of the model, 
and thus reduce the uncertainties in the extracted value of $g_{\phi NN}$. 
In this connection, current experimental efforts at COSY (see contributions by M. 
Wolke, D. Grzonka and A. Khoukaz at this meeting) are of particular interest.  
In any case, the value extracted here is compatible with 
the OZI value of $g_{\phi NN} \cong - (0.60 \pm 0.15)$. More recent data with an improved 
data analysis from the DISTO collaboration \cite{Disto2} exhibit a nearly isotropic angular 
distribution, corroborating a very small nucleonic current contribution.

\begin{figure}[h]
\begin{center}
\leavevmode
\epsfxsize=10.0cm
\epsfysize=7.0cm
\epsfverbosetrue
\epsffile{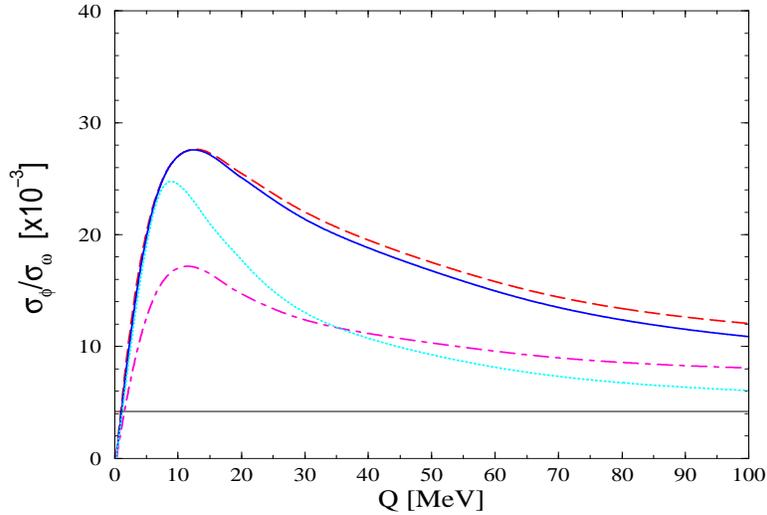}
\end{center}
\caption{Total cross section ratio $R_{\phi/\omega} = \sigma_{pp\rightarrow pp\phi} / 
\sigma_{pp\rightarrow pp\omega}$ as a function of excess energy. Different curves 
correspond to the possible sets of parameters as determined in Ref.\protect\cite{Nak2}.
The horizontal line corresponds to the OZI prediction.}   
\label{fig5}
\end{figure}

Fig.~\ref{fig5} shows the predicted total cross section ratio $R_{\phi/\omega} =
\sigma_{pp\rightarrow pp\phi} / \sigma_{pp\rightarrow pp\omega}$ as a function of excess energy.  
For low excess energies, the predicted ratio is about 4 to 7 times that of the OZI estimate.
At higher energies, the enhancement over the OZI estimate decreases to a factor of 3 or so.
What then is the origin of this enhancement over the OZI estimate as predicted by our model? 
One source of the enhancement is in the mesonic current. As discussed in detail in 
Ref.~\cite{Nak2}, the violation of the OZI rule at the $\phi\rho\pi$ vertex had to be 
introduced in order to achieve a simultaneous and consistent description of the then available
data on the reactions $pp\rightarrow pp\omega$ and $pp\rightarrow pp\phi$. This explicit
OZI violation in terms of the $\phi\rho\pi$ and $\omega\rho\pi$ coupling constants used suggests 
an enhancement of around 3 in the cross section ratio. With regard to the nucleonic current,
the employed $\omega NN$ and $\phi NN$ coupling constants lead to results that exceed the OZI
value only in one case: namely for the parameter set with $g_{\phi NN}= -0.9$ \cite{Nak2}. The
corresponding enhancement factor for the cross section ratio amounts to about 2. 
It is then
evident from the above consideration that the cross section ratios resulting from the model
calculation differ significantly from those values implied by the employed coupling
constants. Obviously, dynamical effects such as interferences, etc., play a rather important
role here and can lead to a fairly large deviations from the OZI prediction within a
conventional picture, i.e., without introducing any ``exotic'' mechanisms. Consequently, one 
should be very cautious in drawing direct conclusions regarding the strangeness content in the 
nucleon from such cross section ratios. The behavior of the cross section ratio as the 
excess energy approaches zero is due to the finite width of the $\omega$, which prevents the
$\omega$ meson production cross section from decreasing rapidly, as it does in the case of
$\phi$ meson.

Evidently, the enhancement of a factor of 3 or so over the OZI estimate
at higher excess energies in Fig.~\ref{fig5} is much smaller than the 
enhancement over the OZI prediction of about a factor 10 found 
by the DISTO collaboration \cite{Disto1} at an excess energy 
$Q \approx$ 80\ MeV in $\phi$ meson production. However, it is important to
realize that their measurement was done at a fixed incident beam
energy of $2.85\ GeV$, and therefore the corresponding 
excess energy of the produced $\omega$ is already $319\ MeV$. Though
corrections for the differences in the available phase space were 
obviously applied when extracting the above result, there are other
effects that may influence the ratio, such as the energy dependence of the
production amplitude, the onset of higher partial waves, etc.,
that cannot be corrected for easily. It is therefore possible that
the actual deviation in the value of $R_{\phi/\omega}$ from the OZI prediction 
is also smaller. Thus, it would be interesting to perform a measurement of 
the cross section ratio at the same (or at least similar) excess energies. 
From the theoretical side, the newer data on $\phi$ production by the DISTO 
collaboration \cite{Disto2} and, especially, the new data on $\omega$ production 
from COSY, which will become available soon \cite{Birkman} 
(see also the contribution by A. Khoukaz at this meeting), should already impose 
much more stringent constraints on the parameters of our model and help address 
better the problem of OZI violation and related issues in $NN$ collisions. In fact,
these new data seem to indicate that we overpredict the $\omega$ meson production cross 
section above $Q \sim 30\ MeV$. This has a direct implication on the cross section ratio
$R_{\phi/\omega}$ at higher excess energies as discussed above.

\subsection{$NN\rightarrow NN\eta$}

We now turn to the production of pseudoscalar mesons. Here we confine our discussion to the 
production of $\eta$ meson in $NN$ collisions. The production of this meson near the threshold 
energy is of special interest, since the existing data are by far
the most accurate and complete among those for heavy meson production and, consequently, 
they offer a possibility to investigate this reaction in much more detail than any of the other 
heavy meson production reactions. In addition to the total cross sections for the $pp \rightarrow 
pp\eta$ reaction \cite{spes3eta1,spes3eta2,celsiuseta,cosy11eta,Calenp}, we have data for
$pn \rightarrow pn\eta$ \cite{Calenn} and $pn \rightarrow d\eta$ \cite{Calenp,Calend}. 
The differential cross section data for the $pp \rightarrow pp\eta$ reaction 
\cite{Calenh} are available as well. Consequently, there is a large number of theoretical 
investigations on these reactions. The production of $\eta$ mesons in 
$NN$ collisions is thought to occur predominantly through the excitation (and de-excitation) of 
the $S_{11}(1535)$ resonance, to which the $\eta$ meson couples strongly. However, the excitation 
mechanism of this resonance is currently an open issue. For example, Batini\'c et al.
\cite{Batinic} (see also the contribution by S. Ceci to this meeting) have found both the 
$\pi$ and $\eta$ exchange as the dominant excitation mechanism. However, they have 
considered only the $pp \rightarrow pp\eta$ reaction. Gedalin et al. \cite{Gedalin} and 
F\"aldt and Wilkin \cite{Wilkin} have considered both the $pp \rightarrow pp\eta$ and 
$pn \rightarrow pn\eta$ reactions. In the analysis of Ref.\cite{Wilkin} the 
$pn \rightarrow d\eta$ reaction was also included. These authors \cite{Gedalin,Wilkin} find the 
$\rho$ exchange to be the dominant excitation mechanism of the $S_{11}(1535)$ resonance. 
In particular, it has been claimed \cite{Wilkin} that $\rho$ meson exchange is important for 
explaining the observed shape of the angular distribution in the $pp \rightarrow pp\eta$ 
reaction. We also mention that, in contrast to the dominant resonance current contribution found 
in Refs.\cite{Batinic,Gedalin,Wilkin}, in a recent calculation of the $pp\rightarrow pp\eta$ 
reaction by Pe\~na et al. \cite{Pena}, it is found that the dominant 
contribution arises not from the $S_{11}(1535)$ resonance current, but from what they refer to as 
the short range amplitude. In our language this corresponds to the shorter range part of the 
nucleonic current. Here we shall report on yet another possible scenario that
reproduces both the $pp \rightarrow pp\eta$ and $pn \rightarrow pn\eta$ reactions and discuss
a possibility to disentangle these reaction mechanisms. 

Although here we shall confine ourselves to the problem just mentioned, the description of 
$\eta$ meson production in $NN$ collisions presents other interesting aspects. For example, 
the $\eta$ meson interacts much more strongly with the nucleon than do mesons like the pion 
so that not only the $NN$ FSI, but also the $\eta N$ FSI is likely to play an important role, 
thereby offering an excellent opportunity to learn about the $\eta N$ interaction 
at low energies. In fact, the near-threshold energy dependence of the observed total 
cross section for $\eta$ meson production differs from that of pion and $\eta^\prime$ production, 
which follow the energy dependence given simply by the available phase-space together with the $NN$ 
FSI. The enhancement of the measured cross section at small excess energies in $\eta$ production 
compared to those in $\pi$ and $\eta^\prime$ production is generally attributed to the 
strong attractive $\eta N$ FSI. 
In addition to all these issues, the theoretical understanding of $\eta$ meson production in $NN$ 
collisions near threshold in free space is also required for investigating the dynamics of the 
$S_{11}(1535)$ resonance in the nuclear medium, possible existence of $\eta NN$ bound states, and 
the possibility of using $\eta$ to reveal the properties of high-density nuclear matter created 
in relativistic heavy-ion collisions.

\begin{figure}[h]
\begin{center}
\leavevmode
\epsfxsize=10.0cm
\epsfysize=7.0cm
\epsfverbosetrue
\epsffile{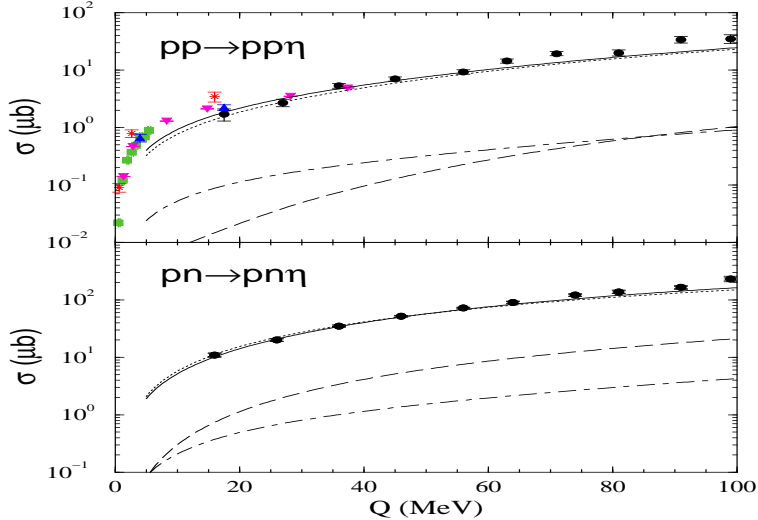}
\end{center}
\caption{Total cross sections for the $pp\rightarrow pp\eta$ (upper panel) and $pn\rightarrow pn\eta$ 
(lower panel) reactions as a function of excess energy. The dashed curves correspond to the nucleonic
current contribution while the dash-dotted curves to the mesonic current contribution; the dotted 
curves represent the resonance current contribution. The solid curves are the total contribution. 
The data are from Refs.\protect\cite{spes3eta1,spes3eta2,celsiuseta,cosy11eta,Calenp,Calenn}.}
\label{fig6}
\end{figure}

\begin{figure}[h]
\begin{center}
\leavevmode
\epsfxsize=10.0cm
\epsfysize=7.0cm
\epsfverbosetrue
\epsffile{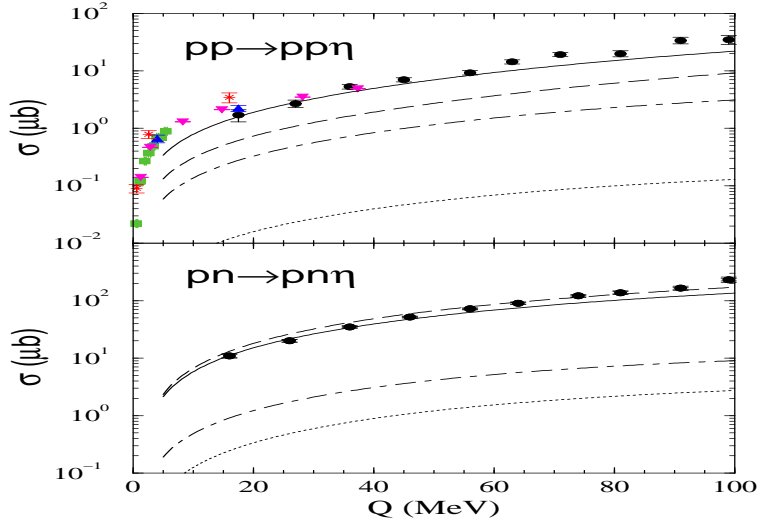}
\end{center}
\caption{Same as Fig.~\ref{fig6}, except that it shows the $S_{11}(1535)$ resonance contribution
only. The dashed curves correspond to the $\pi$ exchange contribution while the dash-dotted 
curves to the $\eta$ exchange contribution; the dotted curves represent the $\rho$ exchange 
contribution. The solid curves show the total contribution.} 
\label{fig7}
\end{figure}

Our model for the $\eta$ production current includes the nucleonic, mesonic and resonance currents.
The mesonic current consists of the $\eta\rho\rho$, $\eta\omega\omega$, and $\eta a_o\pi$ exchange
contributions. The resonance current consists of the $S_{11}(1535)$, $P_{11}(1440)$, and 
$D_{13}(1520)$ resonances excited via $\pi$, $\eta$, $\rho$ and $\omega$ exchange. For the 
$NN$ FSI, we use the Bonn interaction \cite{MHE87}. For the $NN$ ISI, we consider only the 
on-shell interaction obtained from Ref.\cite{SAID}. The details of the calculation will be reported
elsewhere \cite{Nak5}. The results for the total cross section as a function of excess energy are 
shown in Fig.~\ref{fig6}. 
The 
upper panel shows the results for $pp$ collisions while the lower panel those for $pn$ collisions. 
The dashed curves are the nucleonic current contribution; the dash-dotted curves correspond to the 
mesonic current and the dotted curves to the resonance current. The solid curves are the total 
contribution. As one can see, the total cross sections are dominated by the resonance current, 
and more specifically by the strong $S_{11}(1535)$ resonance (see Fig.~\ref{fig7}). 
Our nucleonic current contributions (dashed curves) are much smaller than the resonance current 
contributions. This is in contradiction to the findings of Ref.\cite{Pena}; there, instead of the 
resonance current, the short range amplitude (shorter range part of the nucleonic current) 
gives a large contribution to the $pp\rightarrow pp\eta$ cross section. For small excess 
energies, our calculation underpredicts the data. As mentioned above, this is usually attributed 
to the $\eta N$ FSI, which is not accounted for in the present model. Note that the results for 
$pn \rightarrow pn\eta$ with excess energy $Q > 50\ MeV$, corresponding to incident 
beam energy larger than $1.3\ GeV$, should be interpreted with caution, as no reliable $NN$ 
phase shift analyses for $T=0$ states exist at present for energies above $1.3\ GeV$ \cite{Igor}. 

In Fig.~\ref{fig7} the $S_{11}(1535)$ resonance contribution to the total cross 
sections is shown (solid curves), together with the contribution from the individual meson exchange 
excitation mechanism. The dashed curves correspond to the $\pi$ exchange, while the dash-dotted curves
correspond to the $\eta$ exchange contribution. The dotted curve is due to $\rho$ exchange. 
Although the $\omega$ exchange is included in the calculation, its contribution is not shown here
separately because it is much smaller than the $\rho$ exchange contribution.   
As can be seen, the dominant contribution is due to the $\pi$ exchange followed by $\eta$ exchange.
The $\rho$ exchange is very small. Several observations are in order here:
\begin{itemize}  
\item[1)]
The major reason for the small $\rho$ exchange contribution, in contrast to the result of 
Refs.\cite{Gedalin,Wilkin}, is that in our model we have not allowed the vector ($\gamma^\mu$) 
coupling in the $\rho NN^*$ vertex for spin-1/2 resonances. Note that for an odd-parity spin-1/2
resonance, there is an overall extra $\gamma^5$ factor. Unlike the case of $\rho NN$ vertex, the 
presence of this coupling in the $\rho NN^*$ vertex leads to a violation of gauge invariance, 
which is especially relevant when used in connection with the VMD. As mentioned in the previous 
section, the $\rho NN^*$ coupling constant in our model is extracted from the radiative decay, 
$N^*\rightarrow \gamma + N$, using the VMD. The simplest way of satisfying the gauge invariance 
constraint is to omit the $\gamma^\mu$ coupling from the $\rho NN^*$ vertex, which we have done 
in the present work. Note that the tensor ($\sigma^{\mu\nu}$) coupling is free of this problem. 
A direct consequence of omitting the vector coupling in the $\rho NN^*$ vertex is a very small 
$\rho$ exchange contribution to the cross section as shown in Fig.~\ref{fig7}. Note that in 
Ref.\cite{Gedalin} the $\rho NS(1535)$ vertex is given by the $\gamma^5\gamma^\mu$ coupling. An 
alternative to avoid the gauge invariance problem while keeping the $\gamma^\mu$ term is to use
a vertex of the form $\Gamma^\pm [\gamma^\mu q^2  - (m_{N^*}\mp m_N)q^\mu]$ \cite{Riska}, where
$\Gamma^- \equiv \gamma^5$ and $\Gamma^+ \equiv 1$. In fact, Pe\~na et al. \cite{Pena} 
have used a modified version of this vertex in conjunction with the coupling constant 
determined from a quark model \cite{Riska}. They 
found a non-negligible contribution of the $\rho$ exchange to the excitation of $S_{11}(1535)$ in 
$pp\rightarrow pp\eta$. Relevant experimental information should decide whether the vector coupling 
is required or not in the $\rho NN^*$ vertex. 
\item[2)]
The $\eta$ exchange contribution is relatively large in the present calculation. In fact,
in the case of $pp \rightarrow pp\eta$ its contribution to the cross section is about half of that 
due to the $\pi$ exchange. The $\eta NN$ coupling strength is subject to a rather large uncertainty; 
the value of this coupling constant ranges from $g_{NN\eta} \sim 2.7$ to $g_{NN\eta} \sim 6.4$ 
\cite{Benmer}. The relatively large contribution of $\eta$ here result from using the $\eta NN$ 
coupling constant of $g_{\eta NN} = 6.14$, as used in the construction of the Bonn $NN$ interaction 
\cite{MHE87}. This value is close to the upper limit. However, the $\eta$ meson exchange in the Bonn 
potential \cite{MHE87} might just be an exchange of a $(J^P,T)=(0^-,0)$ quantum number and not of a 
genuine $\eta$ meson. Consequently, the contribution from $\eta$ exchange is subject to this 
uncertainty in the coupling constant. Anyway, in the present calculation for $pp \rightarrow pp\eta$, the 
$\eta$ exchange interferes constructively with the dominant $\pi$ exchange contribution, yielding the 
total contribution as shown by the solid line in Fig.~\ref{fig7}. For $pn \rightarrow pn\eta$, the 
$\eta$ exchange interferes constructively with the $\pi$ exchange in the $T=1$ channel (as in the case 
of $pp \rightarrow pp\eta$), but destructively in the $T=0$ channel due to the isospin factor $-3$ in 
the $\pi$ exchange amplitude. 
\item[3)]
The correct description of both  
$pp \rightarrow pp\eta$ and $pn \rightarrow pn\eta$ reactions depends not only on different isospin 
factors involved for isoscalar and isovector mesons exchanged but, also on a delicate interplay between
the $NN$ FSI and ISI in each partial wave involved. While the $NN$ FSI enhances the total cross 
section, the $NN$ ISI has an opposite effect (see discussion in 
section II). In this connection, we mention that in Ref.\cite{Wilkin} the reduction 
factor due to the $NN$ ISI is estimated to be about $(0.77)^2=0.59$ due to the $^3P_0$ state and 
$(0.73)^2=0.53$ due to $^1P_1$. In our calculation, however, the corresponding reduction factors
are about $0.19$ and $0.27$ near the threshold energy. This large discrepancy between the results of 
Ref.\cite{Wilkin} and ours is due to the fact that, whereas our reduction factor is given by 
Eq.(\ref{isieff}), the reduction factor in Ref.\cite{Wilkin} is given by $\lambda \equiv \eta_L^2 = 
(e^{-Im(\delta_L)})^2$. We argue that the latter formula is inappropriate to estimate the effect of 
the $NN$ ISI for it does exhibit a pathological feature: namely, when the absorption is 
maximum ($\eta_L=0$), this formula yields $\lambda = 0$, implying the total absence of the $NN$ 
elastic channel and thus not allowing the production reaction to occur. However, scattering theory 
tells us that when the absorption cross section is maximum, the corresponding elastic cross 
section does not vanish, but is 1/4 of the absorption cross 
section. Note that this feature is present in Eq.(\ref{isieff}). Furthermore, the authors of 
Ref.\cite{Wilkin} apparently have identified incorrectly the inelasticity $\eta_L$ with 
$\cos^2(\rho)$, where $\rho$ is one of the two parameters (the other is the phase shift) given in 
Ref.\cite{SAID}. The phase shift parametrization given in Ref.\cite{SAID} differs from the standard 
Stapp parametrization as given by Eq.(\ref{phsft}). It is obvious that with a more appropriate 
estimate of the reduction factor $\lambda$ as given by Eq.(\ref{isieff}) the result of 
Ref.\cite{Wilkin} would underpredict considerably the cross section data.
\end{itemize}

\begin{figure}[h]
\begin{center}
\leavevmode
\epsfxsize=10.0cm
\epsfysize=7.0cm
\epsfverbosetrue
\epsffile{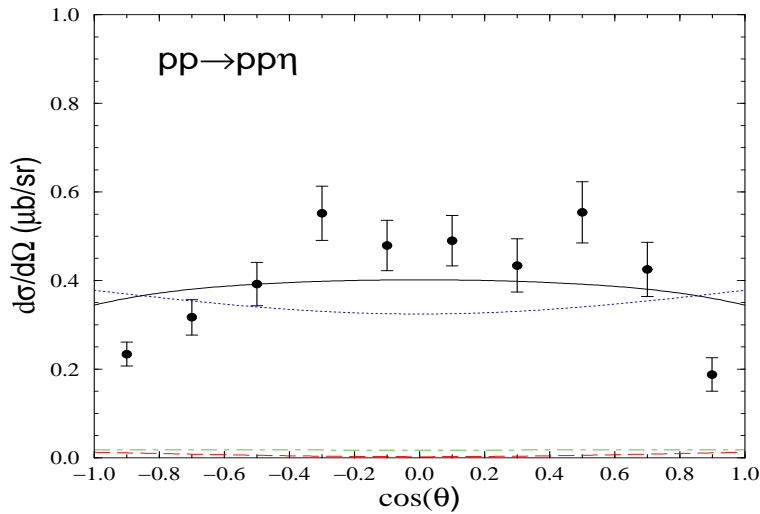}
\end{center}
\caption{Angular distribution of the emitted $\eta$ meson in the c.m. frame of the total
system at an excess energy of $Q=37\ MeV$. The dashed curve corresponds to
the nucleonic current contribution while the dash-dotted curve to the mesonic current 
contribution; the dotted curves represent the resonance current contribution. The solid curve 
show the total contribution. The data are from Ref.\protect\cite{Calenh}.}
\label{fig8}
\end{figure}

Fig.~\ref{fig8} shows the predicted angular distribution of $\eta$ in $pp \rightarrow pp\eta$ at an
excess energy of $Q=37\ MeV$ together with the data of Ref.\cite{Calenh}. 
Again, the resonance 
contribution (dotted curve) dominates the cross section. As pointed out in Ref.\cite{Wilkin}, the 
shape of the angular distribution of the latter contribution bends upwards at the forward and 
backward angles due to the $\pi$ exchange dominance in the $S_{11}(1535)$ resonance contribution.
However, due to an interference with the nucleonic (dashed) and mesonic (dash-dotted) currents, the 
shape of the resulting angular distribution (solid curve) is inverted with respect to that of the 
resonance current contribution alone. As one can see, although the overall magnitude is rather well 
reproduced, the rather strong angular dependence exhibited by the data is not reproduced by our model. 
At this point one might argue that the excitation mechanism of the $S_{11}(1535)$ resonance as given 
by the present model is not correct and that, indeed, the $\rho$ meson exchange should give the 
dominant contribution, as has been claimed in Ref.\cite{Wilkin}. However, judging the level of 
agreements between the predictions of Ref.\cite{Wilkin} and ours with the data, one 
cannot discard the dominance of $\pi$ and $\eta$ exchange in favor of the $\rho$ exchange mechanism 
for exciting the $S_{11}(1535)$ resonance. In this connection, we mention that the new data 
from COSY which will become available soon shows a flat angular distribution \cite{Kilian}.

\begin{figure}[h]
\begin{center}
\leavevmode
\epsfxsize=10.0cm
\epsfysize=7.0cm
\epsfverbosetrue
\epsffile{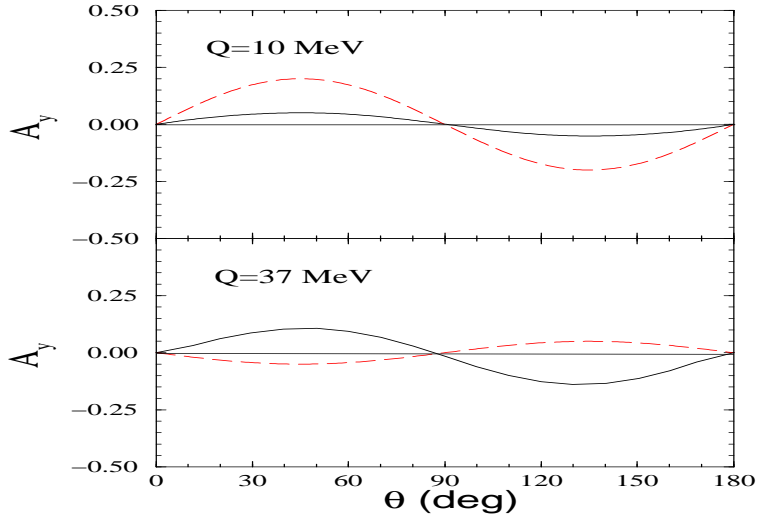}
\end{center}
\caption{Analyzing power for the reaction $pp\rightarrow pp\eta$ as a function of emission-angle
of $\eta$ in the c.m. frame of the total system at an excess energy of $Q=10\ MeV$ (upper
panel) and $Q=37\ MeV$ (lower panel). The dashed curve corresponds to the case of $\rho$ exchange 
dominance according to Ref.\protect\cite{Wilkin}. The solid curve corresponds to the 
present model calculation.}
\label{fig9}
\end{figure}

From the above considerations, we conclude that, at present, the excitation mechanism of the 
$S_{11}(1535)$ resonance in $NN$ collisions is still an open question. Indeed, we have just offered 
a scenario other than the $\rho$ exchange mechanism that reproduces the available data equally well. 
It is therefore of special interest to seek a way to disentangle these possible scenarios. In this
connection, spin observables may potentially help resolve this issue. According to Ref.\cite{Wilkin}, 
the $\rho$ exchange contribution is expected to lead to an analyzing power given by
\begin{equation}
A_y = A_y^{max}\sin(2\theta) \ 
\label{rhoAy}
\end{equation}
in $pp \rightarrow pp\eta$, where $A_y^{max}$ is positive for low excess energies,
peaking at $Q \approx 10\ MeV$ and becoming negative for excess energies $Q > 35\ MeV$. The results 
(dashed curves) at $Q=10\ MeV$ (upper panel) and $Q=37\ MeV$ (lower panel) are shown in Fig.~\ref{fig9}.
The corresponding predictions of the present model are also shown (solid curves). 
The different features exhibited by the two scenarios for the excitation mechanism of 
the $S_{11}(1353)$ is evident. In this connection, the COSY11 effort to measure the analyzing power
in $pp\rightarrow pp\eta$ (see the contribution by P. Winter in this meeting) is of great importance.
We emphasize that these different results should be interpreted with caution. The reason for this is 
that, as mentioned before, the present model accounts for the $NN$ ISI only in the on-shell 
approximation. While this may be a reasonable approximation for calculating cross sections, it may 
introduce rather large uncertainties in the calculated spin observables.

\section{Summary}
The production of heavy mesons in $NN$ collisions has been discussed within the meson exchange model of 
hadronic interactions, paying special attention to the basic dynamics that determine the behavior 
of the cross sections near the threshold energy. Differences in the existing meson exchange models
as well as their limitations have been also discussed. Heavy meson production processes necessarily 
probe the short range dynamics, a domain where we have very limited knowledge so far. In this 
regime even the relevant reaction mechanisms are largely unknown. The theory of heavy meson production 
in $NN$ collisions is still in its early stage of development. Successful description of these 
processes in terms of purely hadronic degrees of freedom calls for correlation of as 
many independent data as possible in a consistent way. The $pp\rightarrow pp\phi$ reaction has 
been discussed as an example of the production of vector mesons in $NN$ collisions.
As for the pseudoscalar meson  production, results for the $\eta$ production in both the $pp$ and 
$pn$ collisions were presented. From these examples, it is clear that not only the  
$pp\rightarrow ppM$, but also the $pn\rightarrow pnM$ and $pn\rightarrow dM$ reactions should be
investigated. Also more exclusive observables than the total cross section such as the spin 
observables should be studied.

{\bf Acknowledgment}
I would like to take this opportunity to thank the organizers of this symposium for
inviting me to present this lecture. I also thank H. Arellano, J. Durso, J. Haidenbauer, 
C. Hanhart, H. Lee, J. Speth, and A. Szczurek who have collaborated in our study of heavy 
meson production reactions at one stage or another. I thank J. Durso, J. Haidenbauer and 
C. Hanhart for a careful reading of this manuscript. 

\vfill \eject

\end{document}